\documentclass{aastex}
\usepackage{myemulateapj5}
\usepackage{lscape}
\begin{document}

\title{A CNO Dichotomy among O2 Giant Spectra\break in the Magellanic Clouds}
\author{Nolan R.\ Walborn}
\affil{Space Telescope Science Institute,\altaffilmark{1} 3700 San Martin Drive, Baltimore, MD 21218} 

\author{Nidia I.\ Morrell\/\altaffilmark{2,3}}
\affil{Las Campanas Observatory, The Carnegie Observatories, Casilla 601,
La Serena, Chile}

\author{Ian D.\ Howarth}
\affil{Department of Physics and Astronomy, University College London,
Gower Street, London WC1E 6BT, UK}

\author{Paul A.\ Crowther\/\altaffilmark{4}}
\affil{Department of Physics and Astronomy, University of Sheffield, 
Hounsfield Road, Sheffield S3 7RH, UK}

\author{Daniel J.\ Lennon}
\affil{Isaac Newton Group, Apartado 321, 38700 Santa Cruz de La Palma,
Canary Islands, Spain}

\author{Philip Massey\/\altaffilmark{2}}
\affil{Lowell Observatory, 1400 West Mars Hill Road, Flagstaff, AZ 86001}

\author{Julia I.\ Arias\/\altaffilmark{2,5}}
\affil{Facultad de Ciencias Astron\'omicas y Geof\'{\i}sicas, Universidad
Nacional de La Plata, Paseo del Bosque, 1900 La Plata, Argentina}
\bigskip
\centerline{\titlelarge To be published in the 20 June 2004 issue of \emph{The Astrophysical Journal}}
\received{14 October 2003}
\accepted{27 February 2004}
\sluginfo
\vfill
\altaffiltext{1}{Operated by the
Association of Universities for Research in Astronomy, Inc., under NASA
contract NAS5-26555.} 
\altaffiltext{2}{Visiting Astronomer, Cerro Tololo 
Inter-American Observatory, National Optical Astronomy Observatory, 
operated by the Association of Universities for Research in Astronomy, Inc., 
under a cooperative agreement with the NSF.}
\altaffiltext{3}{Member of the
Carrera del Investigador Cient\'{\i}fico, CONICET, Argentina; on leave
from Facultad de Ciencias Astron\'omicas y Geof\'{\i}sicas, Universidad
Nacional de La Plata.}
\altaffiltext{4}{Visiting Astronomer, European Southern
Observatory Very Large Telescope; program 70.D-0164.}
\altaffiltext{5}{Fellow of CONICET, Argentina.}

\begin{abstract}
From a survey of the 3400~\AA\ region in the earliest O-type spectra, we 
have found that two of the four O2 giants observed in the Large Magellanic 
Cloud have the O~IV lines there stronger than the N~IV, while the other two 
have the opposite.  A Small Magellanic Cloud counterpart also has the N~IV 
stronger than O~IV.  Inspection of the blue spectra of these stars shows 
that the former pair have weaker N lines in all ionization states (III, IV, 
V) present as well as lines of C~IV $\lambda$4658, while the latter three
have stronger N lines and greater He/H.  Space ultraviolet observations of 
two of the N-strong stars show N~V wind profiles substantially stronger
than those of C~IV, while in the N-weak stars the C~IV features are equal 
to or stronger than the N~V.  The N-strong stars are now reclassified as
ON2~III(f*), newly defining that category.  These characteristics strongly
suggest a larger fraction of processed material in the atmospheres of the
ON2 stars, which we confirm by modeling the optical spectra.  In the context 
of current models, it is in turn implied that the ON2 stars are in a more 
advanced evolutionary state than the others, and/or that they had higher 
initial rotational velocities.  The recent formulation of the effects of 
rotation on massive stellar evolution introduces an additional fundamental 
parameter, which the CNO abundances are in principle able to constrain.  
We present some illustrative comparisons with current Geneva evolutionary 
models for rotating massive stars.  It is possible that these very hot,
nitrogen-rich objects are products of homogeneous evolution.  Our results 
will provide motivation for further physical modeling of the atmospheres and 
evolutionary histories of the most massive hot stars.  
\end{abstract}

\keywords{Magellanic Clouds --- stars: abundances --- stars: early-type --- 
stars: evolution --- stars: fundamental parameters}

\section{Introduction}

Anomalies in the relative intensities of CNO lines in OB spectra have
been described for some time (Walborn 1976, 1988, 2003; and references
therein).  Anticorrelations of N vs.\ C, O and correlations with He/H in 
the expected sense have encouraged interpretations in terms of mixing of 
CNO-cycled material into the atmospheres and winds of massive stars.  An
important point has been the recognition that the morphologically normal
majority of OB supergiants actually have some degree of N and He
enhancements already, while the OBC minority have normal (i.e.,
main-sequence) abundances and the OBN group have more extreme mixing
(Maeder \& Conti 1994; Smith \& Howarth 1994).  The recent development of 
models for massive stellar evolution with rotation (Maeder \& Meynet 2000) 
is an advance toward realism since massive stars often rotate rapidly, 
but another dimension is added to the phase space that must be explored
in order to compare observations with theory.  Fortunately, a new
observational diagnostic is added concurrently, namely the CNO abundances, 
since the degree of mixing also depends on the initial rotation.  Indeed, 
progress has already been made toward understanding apparent discrepancies 
among the HRD locations of some mid-O giants and supergiants in the Small 
Magellanic Cloud (SMC; Walborn et~al.\ 2000), in terms of their CNO abundances 
and inferred initial rotations (Crowther et~al.\ 2002; Hillier et~al.\ 2003).  
Related results for main-sequence stars in the SMC giant H~II region NGC~346 
are reported by Bouret et~al.\ (2003).

Spectral classification of the hottest massive stars is relatively difficult,   
because of their scarcity and the paucity of reliable features in their
optical spectra.  Recently, an increased sample of high-quality data has
supported classification developments for them, including introduction of
the new types O2 and O3.5 (Walborn et~al.\ 2002).  In this system, reliance is 
placed on ratios of the selective emission lines (Walborn 2001) of N~IV and 
N~III, primarily on an empirical basis.  These new spectral types have yet to 
be modeled systematically and calibrated in terms of physical parameters, 
including confirmation that the O2 extension corresponds to higher effective 
temperatures as hypothesized.  The analysis of these spectra is also difficult, 
because of uncertainties in the parameters and the atmospheric physics.  E.g., 
the special behavior of the N~IV and N~III emission lines has to be reproduced 
and interpreted.  Currently, modeling and calibration developments are 
underway to address these issues; we shall present some relevant results here.

Certainly additional spectral features to constrain the classification
and analysis of the hottest spectra would be welcome.  Drissen et~al.\ 
(1995) pointed out certain little-observed N~IV and O~IV lines in the 
3400~\AA\ region of early-O spectra, that warranted further investigation
in a larger sample including longer-wavelength classification standards.  
We have now accomplished such a survey, which will be fully reported by
Morrell et~al.\ (in preparation).  Here we describe our unanticipated
discovery of the first CNO anomalies in optical O2 spectra, which are
relevant to both the classification and modeling of these extreme objects.

\section{Observations}

The far-violet observations of the Large Magellanic Cloud (LMC) stars were 
carried out with the Cassegrain spectrograph at the Cerro Tololo Inter-American 
Observatory (CTIO) 1.5~m telescope during 2002 February~22--27.  A 
600~lines~mm$^{-1}$ grating blazed at 3375~\AA\ in second order combined with 
a Loral 1k CCD provided a 2~pixel resolution of 1.5~\AA\ and wavelength 
coverage from 3270 to 4180~\AA.  The LMC stars discussed here have 
($V$, $B-V$) ranging from (12.28, $-$0.23) for HDE~269810 to (13.68, 
$-$0.01) for LH10-3061 (see Walborn et~al.\ 2002 for detailed stellar 
parameters).  Total exposure times ranging from~1 to 3~hours yielded
signal-to-noise values between 60 and 150 per pixel at 3400~\AA, 
reflecting the low (but relatively high at CTIO) atmospheric transmission 
at that wavelength.  The data for all but HDE~269810 have been smoothed 
by 3~pixels for presentation here, and the radial velocities have been
corrected to rest.  The data reductions and analysis were performed with 
standard IRAF\/\footnote{IRAF is distributed by the National Optical Astronomy 
Observatory.} routines running under Linux at the La~Plata Observatory.  

NGC~346-3 (13.50, $-$0.23) in the SMC was observed on the night of 2003
July~30--31 with the Boller and Chivens spectrograph at the Las Campanas 
Observatory (LCO) Magellan~I (Walter Baade) 6.5~m telescope.  The detector 
was a $2048 \times 515$ pixel Marconi CCD (pixel size 13.5~$\mu$m).
A 600~lines~mm$^{-1}$ grating blazed at 5000~\AA\ provided wavelength 
coverage from 3300 to 6400~\AA\ with a 2~pixel resolution of 3.1~\AA.
The exposure time was 10~min, producing a signal-to-noise of about 50 per
pixel near both ends of the spectrogram but reaching 300 at 5000~\AA.
The data were reduced and analyzed with standard IRAF routines at LCO.

Details of the blue-violet observations shown here can be found in 
Walborn et~al.\ (2002).  The ultraviolet observations discussed correlatively 
are from the {\it Hubble Space Telescope\/} (\emph{HST}) Faint Object Spectrograph
(FOS) and Space Telescope Imaging Spectrograph (STIS); they are fully
described by Walborn et~al.\ (1995) and by Massey et~al.\ (2004), respectively. 

As a check on the quantitative analysis of these moderate-resolution
data, an available high-resolution observation of HDE~269810 was also
analyzed.  This observation was made at the European Southern Observatory 
(ESO) Very Large Telescope (VLT), with the UV-Visual Echelle Spectrograph
(UVES) mounted at the UT2 8~m, on 2002 November~29 under program 70.D-0164.  
The resolving power of this system is 70,000, providing a resolution of 
0.05~\AA\ at the O~IV $\lambda3400$ lines.  Two setups with several channels 
containing EEV and MIT/LL 2k~$\times$~4k CCDs with 15~$\mu$m pixels covered 
wavelengths from the far violet through H$\alpha$.  Exposure times of 1200~s 
yielded S/N per 2~pixel resolution element ranging from 80 at 3450~\AA\ to 
150 at 4700~\AA\ and 120 at 6600~\AA.  The data were reduced by the UVES
pipeline under the Munich Image Data Analysis System (MIDAS).

\section{Results}

A sample of 27 Galactic and Magellanic Cloud early-O stars has been observed 
in the 3400~\AA\ region; in most of them, the N~IV and O~IV features have 
comparable intensities and display systematic trends with spectral type and 
luminosity class, which will be discussed by Morrell et~al.\ (in preparation).  
Surprisingly, however, the four LMC O2~III(f*) spectra in the sample separate 
into two groups, with O~IV substantially stronger than N~IV in one pair and 
the opposite in the other, while the SMC object is similar to the latter LMC
pair, as shown in Figure~1.  This unexpected
result prompted us to re-examine in a new light the blue-violet spectra of 
the same five stars shown by Walborn et~al.\ (2002) and reproduced in
Figure~2.  It was immediately clear that the systematic differences between 
HDE~269810 and Sk~$-$68$^{\circ}$~137 on the one hand, and LH64$-$16, 
LH10$-$3061, and NGC~346-3 on the other, can be described consistently in terms 
of different relative line strengths of N vs.~C.  In particular, the first 
pair have weaker N~III and N~IV emission lines as well as N~V absorption
lines, along with the presence of C~IV $\lambda4658$, while all of the 
N~lines are stronger in the latter group (except for the N~V in NGC~346$-$3).  
Moreover, the He/H line ratios are seen to be larger in the latter group as 
well (although unclear in the SMC object due to nebular hydrogen emission 
and the lower resolution in the far violet).  Presumably the O~IV
$\lambda4632$, which is also prominent in the first pair, is blended with
N~III $\lambda4634$ in the other stars.  The morphology strongly excludes
ionization effects as an explanation and equally strongly suggests N/(C,O)
and He/H abundance differences as the cause.   

Striking support for this conjecture is provided by the ultraviolet
stellar-wind spectra of these stars.  Those of HDE~269810,
Sk~$-$68$^{\circ}$~137, and NGC~346$-$3 were discussed by Walborn et~al.\ (1995) 
and are reproduced in Figure~3, along with that of LH64-16 obtained and 
analyzed by Massey et~al.\ (2004).  In HDE~269810, the C~IV $\lambda1550$ wind
profile is comparable to that of N~V $\lambda1240$ and in
Sk~$-$68$^{\circ}$~137 the C~IV is actually somewhat stronger than the
N~V, while in LH64$-$16 N~V is stronger and in NGC~346-3 much stronger than C~IV. 
The UV spectrum of LH10$-$3061 has not yet been observed to our knowledge.

Both the optical and UV spectra of LH64$-$16 and NGC~346-3, as well as LH10$-$3061 
in the optical, relative to HDE~269810 and Sk~$-$68$^{\circ}$~137, are
consistent with the characteristics of the ON class (Walborn 1976, 1988,
2003).  Hence, the first three spectra are hereby reclassified as ON2~III(f*).
This class has been interpreted in terms of mixing of CNO-cycled material
into the atmospheres and winds of the stars (e.g., Maeder \& Conti 1994).
The presence of these effects at the earliest spectral types is a new
development that may further challenge the models, however, as discussed below. 

Walborn et al. (2000) had already classified NGC~346-3 as ON3~III(f*), 
purely on the basis of the very large N~V/C~IV wind-profile ratio in its 
UV spectrum.  The ON designation was not retained when this star was moved 
into the O2 class by Walborn et~al.\ (2002), partly because of concerns about 
a pure UV classification criterion and about the possibility that all such 
spectra might have similarly large N/C, but obvious only in the SMC object 
because the lower systemic metallicity allowed the very sensitive C~IV feature 
to desaturate.  However, LH64$-$16 demonstrates that the latter concern is not 
valid, since the C~IV can be weaker than N~V at LMC metallicity.  Thus, the
proper classification of NGC~346$-$3 is also ON2~III(f*), as confirmed here
by the 3400~\AA\ observations.  Haser et~al.\ (1998) derived a very large N/C 
abundance ratio in NGC~346-3, but normal in Sk~$-$68$^{\circ}$~137, while Puls 
et~al.\ (1996) gave normal He/H in both HDE~269810 and Sk~$-$68$^{\circ}$~137, 
consistent with the present classifications and interpretations.  Bouret
et~al.\ (2003) have also derived an enhanced N abundance in NGC~346-3, as
further discussed below.

\section{Spectroscopic Modeling}

In order to gain physical insight into the CNO line-strength differences 
outlined above, we have undertaken spectroscopic modeling of our blue and 
violet spectrograms.  Example fits to the spectra of two stars are shown
in Figure~4.

\subsection{Techniques}

Recent developments in model-atmosphere studies of hot stars routinely
permit sophisticated metal-line blanketing in non-LTE to be considered,
under either plane-parallel geometry (e.g., TLUSTY: Lanz \& Hubeny 2003)
or spherical geometry (e.g., CMFGEN: Hillier \& Miller 1998).  Bouret et al.
(2003) have demonstrated the excellent consistency in synthetic spectra 
achieved between TLUSTY and CMFGEN in a combined UV and optical study of 
NGC~346-3, ON2~III(f*).  The former technique provides a more comprehensive 
treatment of photospheric metal-line blanketing, while the latter offers 
insights into wind properties. 

In the present application, we have utilized CMFGEN, the current version
of which is discussed by Hillier et al. (2003).  The code solves the radiative
transfer equation in the co-moving frame, under the additional constraint of 
statistical equilibrium.  The model used here is similar to that applied by 
Bouret et al. (2003), except that some important additional ions are 
considered.  In total, 27 ions of H, He, C, N, O, Si, P, S, Ar, Fe, and Ni
are included, comprising a total of 2000 individual levels (grouped into
800 superlevels), with a full array of 17,000 bound-bound transitions. The 
temperature structure is determined by radiative equilibrium.  CMFGEN does 
not solve the momentum equation, so that a density or velocity structure is
required.  For the supersonic part, the velocity is parameterized with
a classical $\beta$-type law (with $\beta = 1$), which is connected to a
hydrostatic density structure at depth, such that the velocity and velocity 
gradient match at the interface.  The subsonic velocity structure is set by 
the corresponding fully line-blanketed TLUSTY (v.200) model with 
$\log g = 4.0$ (cgs).

We have assumed a depth-independent Doppler profile for all lines when
solving for the atmospheric structure in the co-moving frame, while in the 
final calculation of the emergent spectrum in the observer's frame, we have 
adopted a radially-dependent turbulence, which reflects the effect of shocks 
due to wind instabilities.  The minimum turbulence, $\xi$, is set from the 
core strengths of H$\gamma$ and He~II $\lambda$4541.  Incoherent electron 
scattering and Stark broadening for hydrogen and helium lines are adopted.  
We convolve our synthetic spectrum with a rotational broadening profile.  
Because of the intermediate dispersion of our main observations, instrumental 
effects prevent a reliable determination of $v \sin i$, so that 100~km~s$^{-1}$ 
has been adopted in each case, in accord with previous results for HDE~269810, 
Sk~$-$68$^{\circ}$~137, and NGC~346-3 (Puls et~al.\ 1996).
   
Temperatures of O stars are generally derived from spectroscopic fits to lines 
of He~I-II (e.g., Conti 1973; Crowther et~al.\ 2002).  Since He~I is absent or 
extremely weak in O2 stars, we rely instead on the N~IV$-$V ionization balance, 
i.e., N~IV $\lambda\lambda$3479$-$85, 4058 and N~V $\lambda\lambda$4604$-$20.  
Of course, the nitrogen abundance of an individual star is not known 
{\it a priori\/}, so that it is also a free parameter.  With the stellar 
temperature set, we then derive the stellar radius from the absolute visual 
magnitude, for which we adopt values presented by Walborn et~al.\ (2002).
Since LH10$-$3061 has not been observed in the UV, we adopt a nominal wind 
terminal velocity of 3200 km~s$^{-1}$ for it, following prior UV analysis of 
Sk~$-$68$^{\circ}$~137 (Prinja \& Crowther 1998); sources of terminal
velocities for the other stars are given in notes to Table~1.  Lacking the 
usual H$\alpha$ wind diagnostic for most of the stars, we estimate mass-loss 
rates from the strength of He~II $\lambda$4686, although we recognize that 
as a less reliable indicator.
 
With regard to the helium abundance, generally He/H~=~0.1 (by number)
is adopted unless noted otherwise.  Potential nebular contamination of 
Balmer cores generally prevents accurate determinations of He/H from 
intermediate-dispersion observations of O~stars.  For NGC~346-3,  
inspection of a high-resolution AAT/UCLES observation (Walborn et~al.\ 2000) 
supports He/H~$\sim 0.1$.  In contrast, the CTIO 4~m observation of LH64$-$16 
(Massey, Waterhouse, \& DeGioia-Eastwood 2000) analyzed here suggests a 
rather helium-rich chemistry, with He/H~$\sim 0.25$.  This same
observation of LH64$-$16 has been modeled independently by Massey et~al.\ 
(2004), who find an even larger He/H abundance ratio.

We estimate carbon and oxygen abundances from C~IV $\lambda$4658 and O~IV 
$\lambda\lambda$3348$-$49, 3381$-$3410, respectively, while all remaining 
elements are set to 0.4~$Z_{\odot}$ (LMC) or 0.2~$Z_{\odot}$ (SMC), following, 
e.g., Russell \& Dopita (1992).  Solar abundances are taken from Grevesse \& 
Sauval (1998; log(C/H)~+~12~=~8.52, log(N/H)~+~12~=~7.92), except that the 
oxygen abundance has recently been revised downward to log(O/H)~+~12~=~8.66 
(Apslund 2003).

\subsection{Results}

Early O giants in the Magellanic Clouds have previously been studied
quantitatively by Puls et~al.\ (1996), who derived stellar temperatures of 
55, 60, and 60~kK for NGC~346$-$3, Sk~$-$68$^{\circ}$~137, and HDE~269810, 
respectively, from plane-parallel photospheric techniques, which neglected 
the effects of line blanketing.  As with other recent redeterminations of 
O-star temperature scales (e.g., Martins, Schaerer, \& Hillier 2002;
Crowther et~al.\ 2002), we find 2.5--5~kK lower temperatures.  Indeed, 
our results closely mimic those of Bouret et~al.\ (2003) for NGC~346$-$3,
with stellar temperatures in the range $52.5 \leq T_{\rm eff} \leq 55$~kK 
and luminosities in the range $8 \times 10^{5} \leq \log(L/L_{\odot}) \leq 
2.2 \times 10^{6}$, plus unclumped mass-loss rates of 1--$2 \times 10^{-6}\  
M_{\odot}$~yr$^{-1}$ (Table~1). 

More important in the present context are elemental abundances (Table~1). 
For the LMC O2 giants, we find only modest increases in the nitrogen 
abundances from the ISM ratios.  In contrast, large N/C and N/O ratios are 
obtained for the LMC ON2 giants, suggesting partially processed material at 
the stellar surfaces.  For NGC~346-3 in the SMC, we confirm processed CNO 
abundances as derived by Bouret et~al.\ (2003).  Haser et~al.\ (1998) previously 
estimated N~$\geq$~0.4~N$_{\odot}$ and C~$\leq$~0.02~C$_{\odot}$ for this 
star from UV spectral fits.  Relative to their corresponding ISM ratios,
we find that N/C is enhanced in the atmospheres of LH64$-$16, LH10$-$3061, and 
NGC~346$-$3 by factors of 46, 46, and 52, and N/O by factors of 14, 23, and 
12, respectively.  These factors can be compared with those of order 100 
derived by Crowther et~al.\ (2002) and Hillier et~al.\ (2003) for later-type 
O~supergiants in the Magellanic Clouds.

\subsection{Uncertainties}

In order to estimate the uncertainties of our abundance results, we have
undertaken additional modeling of two kinds.  First, the fits to the data
discussed above have been re-examined, to determine the range of
parameters, if any, that might yield comparable or perhaps even improved
fits.  For example, in the models upon which the results of Table~1 are
based, the N~IV $\lambda4058$ emission line was well matched while the N~V 
absorption lines were somewhat underpredicted.  It was found that
increasing the $T_{\rm eff}$ of the two ON2 stars, LH64$-$16 and LH10$-$3061, to
56~kK improves the fits to the N~V lines but somewhat overpredicts the
N~IV emission, as shown in Figure~4.  Simultaneously, the fits to the
very weak C~IV $\lambda4658$ emission line, the sole diagnostic of the
carbon abundances in these data, were re-examined, with the result that 
C/C$_{\odot}$ in the ON2 stars could be reduced to as low as 0.02.  The
only other possible changes indicated by this further modeling are small
increases in O/O$_{\odot}$ for the two LMC O2 stars, HDE~269810 and
Sk~$-$68$^{\circ}$~137, to 0.6 and 0.7, respectively.  In particular, no
changes to the derived N~abundances are indicated.  To summarize, as
best we can determine from the intermediate-dispersion data alone, the
uncertainties in $T_{\rm eff}$ are $\pm1$~kK, in the carbon abundances a
factor of 2.5, and in the oxygen abundances a factor of 1.5.  One could
infer that the uncertainty in the nitrogen abundances is less than a
factor of 1.5. 

A second, more powerful comparison is provided by the available VLT/UVES
high-resolution spectrogram of HDE~269810, with wavelength coverage extending 
into the yellow-red.  Not only are the multiple, weak O~IV lines well
resolved and defined in these data, additionally including
$\lambda\lambda$3560$-$63 and 4632, but the strong C~IV $\lambda\lambda5801$, 
5812 emission lines are available to better constrain the carbon
abundance, and N~IV $\lambda6381$ is also available.  
The good fit obtained with the same models used to analyze the 
intermediate-dispersion data is shown in Figure~5.  The only significant
discrepancies are in the O~IV $\lambda\lambda$3404$-$12$-$14 triplet, which is
predicted in emission by models this hot but observed in absorption, and
in higher members of the He~II Pickering series, which remain overpredicted.  
The improved fit to hydrogen and other helium lines is noteworthy; the 
discrepancies at the lower dispersions are believed to be caused primarily by 
line-broadening and resolution matching between the models and data.  It is 
reassuring that the mass-loss rate based also on the H$\alpha$ profile here 
agrees well with that determined previously from He~II $\lambda4686$ alone.
The stellar parameters that differ in the high-resolution analysis are 
$T_{\rm eff}$, 55~kK; $\log L/L_{\odot}$, 6.38; $\log \dot{M} 
(M_{\odot}$ yr$^{-1}$), $-$5.5; and He/H, 0.07; comparison with the
respective values in Table~1 indicates the reliability of that analysis.  
The derived C, N, O abundances relative to solar are 0.05, 0.6, and 0.8, 
respectively; i.e., the C abundance is smaller by a factor of 4, while the 
N, O abundances are larger by respective factors of 1.5 and 2.  It follows 
from these results that HDE~269810 already has a an appreciable N/C 
enhancement over the LMC ISM, by a factor of 23.

The uncertainties in the solar CNO abundances given by Grevesse \& Sauval
(1998) are $\pm0.06$~dex, or a factor of 1.15.  The corresponding Magellanic 
Cloud ISM abundance uncertainties from Russell \& Dopita (1992) are 
0.13--0.18~dex for C, 0.15--0.20~dex for N, and 0.06-0.10~dex for O, or 
average factors of 1.4, 1.5, and 1.2, respectively.  Thus, the uncertainties 
in the ratios are dominated by the stellar abundances, which we have 
estimated above to be factors of 2.5--4 for C, and 1.5--2 for N and O.  
The worst case is the N/C ratio, which in principle could be uncertain by
a factor of~6.  However, such a factor cannot apply randomly to the
results, since it could invert the differential spectral morphology,
which is not reasonable; on the other hand, a systematic effect of that
magnitude cannot be ruled out with the present data.  It is interesting that
all of the C~redeterminations are lower, while the N and O are somewhat
higher.  Thus, the N/C ratios may be even larger than given in Table~1.
Clearly it is highly desirable to reanalyze all of these stars with
high-resolution data, at the same time determining their mass-loss rates
from H$\alpha$ and UV data, to refine the quantitative results.

\section{Discussion: Evolutionary Interpretations}

Curiously, the luminosities, and hence the masses, estimated by Walborn et~al.\ 
(2002) for the O2~III(f*) stars HDE~269810 (150~$M_{\odot}$, average of 
two methods) and Sk~$-$68$^{\circ}$~137 (100~$M_{\odot}$) are considerably
higher than those for the ON2~III(f*) stars LH64$-$16 (62~$M_{\odot}$),
LH10$-$3061 (81~$M_{\odot}$), and NGC~346$-$3 (82~$M_{\odot}$).  The
opposite relationship might have been expected, since both mass-loss
rates and mixing increase with mass (Meynet \& Maeder 2000).  On the
other hand, perhaps the longer evolutionary timescales at the lower
masses favor the observability of mixing effects.  

Until recently, these abundance results would have been interpreted simply 
in terms of more advanced evolutionary ages for the ON2 than the O2
objects, which remains a possibility.  (Of course, mass transfer in a binary 
system is an alternative mechanism likely responsible for at least some 
members of the ON class---see Bolton \& Rogers 1978; Levato et~al.\ 1988;
Walborn \& Howarth 2000; Lennon 2003).  In the context of evolutionary models 
with rotation for massive stars (Maeder \& Meynet 2000; Meynet \& Maeder 
2000), however, the degree of mixing of processed material to the surface at 
a given evolutionary age is a function of the initial rotational velocity. 
Indeed, CNO abundances in massive stars now become critical diagnostics 
of their initial parameters and evolutionary histories.  For a given
metallicity, the observed $T_{\rm eff}$, $L$, and CNO abundances in principle 
enable derivation of the corresponding initial mass, current age, and initial
rotational velocity.  Complete grids of evolutionary tracks for the most
massive stars, at Magellanic Cloud metallicities and with a range of 
rotational velocities, are currently under development in Geneva.
Meanwhile, we display illustrative comparisons in Figures~6--8, courtesy
of G.~Meynet and A.~Maeder.

Figure~6 is a theoretical HR Diagram in which the locations of the stars
discussed here are compared with Geneva evolutionary tracks for initial
masses of 60 and 120~$M_{\odot}$, at both LMC and SMC metallicities (mass
fractions Z~=~0.008 and 0.004, respectively), and with an initial equatorial
rotational velocity of 300~km~s$^{-1}$.  All of the stars lie near the
zero-age main sequence (ZAMS), which is consistent with an early evolutionary 
stage for the two objects with normal spectra; the present spectroscopic 
analysis again indicates higher masses for them than for the ON2 objects.
However, such a location is surprising for the latter, highly mixed 
stars.  Their initial masses still place them above the Humphreys-Davidson
Limit, so they cannot be interpreted as post-red supergiants evolving
blueward.  A more relevant possibility may be that of ``homogeneous'' 
evolution, along tracks that rise and curve blueward from the ZAMS in the 
HRD, which is predicted in the cases of strong mixing and/or very high 
initial rotational velocities (Langer \& El~Eid 1986; Maeder 1987; Langer 
1992; Meynet \& Maeder 2000).  The 60~$M_{\odot}$ homogeneous track at
solar metallicity (Z~=~0.020) with an initial rotational velocity of
400~km~s$^{-1}$ from the last reference is also reproduced in Fig.~6;
it trends near the locations of the ON2~III(f*) points.  Indeed, 
Bouret et al. (2003) have concluded that such a track is required to 
understand NGC~346-3 as coeval with the other early O stars in the cluster.  

Figures~7 and 8 compare our derived N/C and N/O ratios, respectively, relative 
to their corresponding interstellar media, taken at face value from Table~1, 
with the predictions of the same Geneva models.  The N/C ratios in the 
ON2~III(f*) stars are the largest reported to date for near-main-sequence 
objects and are substantially larger than the standard model predictions, 
suggesting even more rapid mixing and perhaps providing additional support for 
homogeneous evolution.  In contrast, the N/O ratios are compatible with the 
standard model predictions and indicate early operation of the ON cycle, as 
expected in only the most massive stars.  

On the other hand, the even larger N enhancements derived for later-type 
O supergiants in the Magellanic Clouds by Crowther et~al.\ (2002) and 
Hillier et~al.\ (2003) may imply that the ON2~III(f*) stars are simply 
predecessors of those objects along redward evolutionary tracks.  Again,
much stronger mixing effects near the main sequence than predicted by 
current models, at least for N/C, are then required by our observations.  
Further analysis of these phenomena, with improved data and relative to the 
latest models for rotating massive stars, will clarify these issues and 
provide an improved understanding of their post-main sequence evolution.

\acknowledgments
We are very grateful to Georges Meynet and Andr\'e Maeder for their
substantial contributions to the discussion of these results, including
Figures~6--8.  We also thank the STScI Director's Discretionary Research 
Fund for publication support.  PM's participation was supported by NASA
through grant GO-8633 from the Space Telescope Science Institute, which is 
operated by AURA, Inc., under contract NAS5-26555.  We thank an anonymous
referee for useful recommendations.

\bigskip
\begin{center}
\begin{tabular}{lcccccc}
\tableline\noalign{\smallskip}
Star               & HDE~269810 & Sk$-68^{\circ}$ 137 & LH64--16 & LH10--3061 & 
NGC~346--3 \\
Sp Type            & O2 III(f*) &  O2 III(f*)  & ON2 III(f*) &  ON2 III(f*) & 
ON2 III(f*) \\[2pt]
\tableline\noalign{\smallskip} 
$T_{\rm eff}$ (kK) &    \llap{5}2.5    &    \llap{5}5~\phn    &   \llap{5}5~\phn    &  \llap{5}5~\phn  &    \llap{5}2.5 \\
$\log L (L_{\odot}$)&   6.3\rlap{4}    &    6.1\rlap{9}  &   5.9\rlap{0}  &  6.0\rlap{8}  &     6.0\rlap{7} \\
$\log \dot{M} (M_{\odot}$ yr$^{-1}$) & \llap{$-$}5.6 & \llap{$-$}5.7 &  \llap{$-$}5.8
& \llap{$-$}6.0 &   \llap{$-$}5.8    \\
$v_{\infty}$(km s$^{-1}$) & \llap{375}0$^{a}$  & \llap{320}0$^{b}$ & \llap{325}0$^{c}$   & 
\llap{320}0:$^{b}$  & \llap{280}0$^{d}$         \\
$M_{V}$ (mag)$^{e}$      &  \llap{$-$}6.6    & \llap{$-$}6.1 & \llap{$-$}5.4 & \llap{$-$}5.8  & \llap{$-$}5.9 \\
$\xi$ (km s$^{-1})$      & \llap{1}0~\phn         & \llap{1}5~\phn   & \llap{1}0~\phn   & \llap{1}5~\phn     & \llap{2}5~\phn   \\
He/H     & \llap{$\leq$}0.1 & 0.1\rlap{:} & 0.2\rlap{5} & 0.1\rlap{:} & 0.1 \\
\\
C/C$_{\odot}$ (ISM)  &   0.2 (0.33)   & 0.2 (0.33)  &  0.05 (0.33)
& 0.05 (0.33)  & 0.025 (0.16)\\
N/N$_{\odot}$ (ISM)  &   0.4 (0.17)   & 0.3 (0.17)  &  1.2\phn\phn(0.17)
& 1.2\phn\phn(0.17)  & 0.4\phn\phn\phn(0.05)\\
O/O$_{\odot}$ (ISM)  &   0.4 (0.49)  &  0.6 (0.49)  &  0.25 (0.49)
& 0.15 (0.49)   &  0.15\phn\phn(0.23) \\
\\
$\frac{\rm N/C}{\rm (N/C)_{\odot}}$ (ISM) &  2.0 (0.52)  & 1.5 (0.52)   &
\llap{2}4\phn\phn\phn(0.52) & \llap{2}4\phn\phn (0.52)    & \llap{1}6\phn\phn\phn(0.31) \\
$\frac{\rm N/O}{\rm (N/O)_{\odot}}$ (ISM) &  1.0 (0.35)  & 0.5 (0.35)   &
4.8\phn\phn(0.35)  & 8.0 (0.35)    & 2.7\phn\phn(0.22) \\
\\
$\frac{\rm N/C}{\rm (N/C)_{\rm ISM}}$  &  3.8  & 2.9  & \llap{4}6~\phn & \llap{4}6~\phn  & \llap{5}2~\phn \\
$\frac{\rm N/O}{\rm (N/O)_{\rm ISM}}$  &  2.9  & 1.4  & \llap{1}4~\phn & \llap{2}3~\phn  & \llap{1}2~\phn \\
\noalign{\smallskip}
\tableline
\noalign{\smallskip}
\multicolumn{6}{l}{(a) Walborn et al. 1995; (b) Prinja \& Crowther 1998; 
(c) Massey et al. 2004;}\\ 
\multicolumn{6}{l}{(d) Bouret et al. 2003; (e) Walborn et al. 2002}\\
\end{tabular}
\end{center}
\newpage

\begin{figure}
\centering
\includegraphics[width=.9\textwidth]{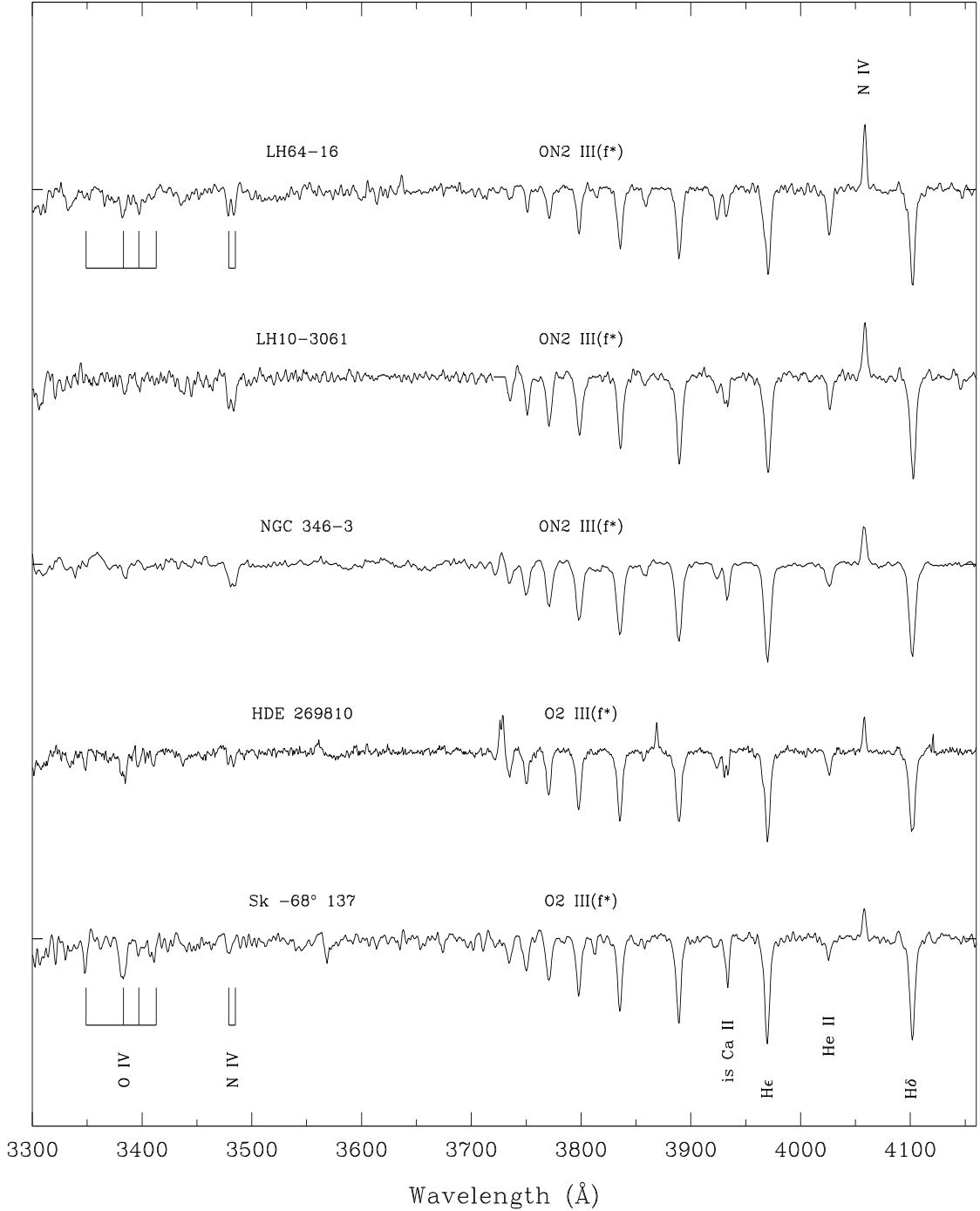}
\caption{
Rectified linear-intensity, far-violet spectrograms of five O2 giants in the 
Magellanic Clouds, separated by 0.5 continuum units.  The absorption lines 
identified below are O~IV $\lambda\lambda$3348$-$49, 3381$-$85, 3397, 3410$-$14; 
N~IV $\lambda\lambda$3479$-$85 (the latter a blend of two lines); interstellar 
Ca~II $\lambda$3933; H$\epsilon$ $\lambda$3970; He~II $\lambda$4026; and 
H$\delta$ $\lambda$4101.  The N~IV $\lambda$4058 emission line is identified 
above (and the O~IV, N~IV absorption-line brackets are replicated).  In the
spectrogram of LH10$-$3061, an oversubtracted [O~II] $\lambda$3727 nebular
emission doublet has been patched out.  Note the different relative
strengths of the O~IV and N~IV features between the ON2 and O2 classes.
\label{fig1}}
\end{figure}

\begin{figure}
\centering
\includegraphics[width=.9\textwidth]{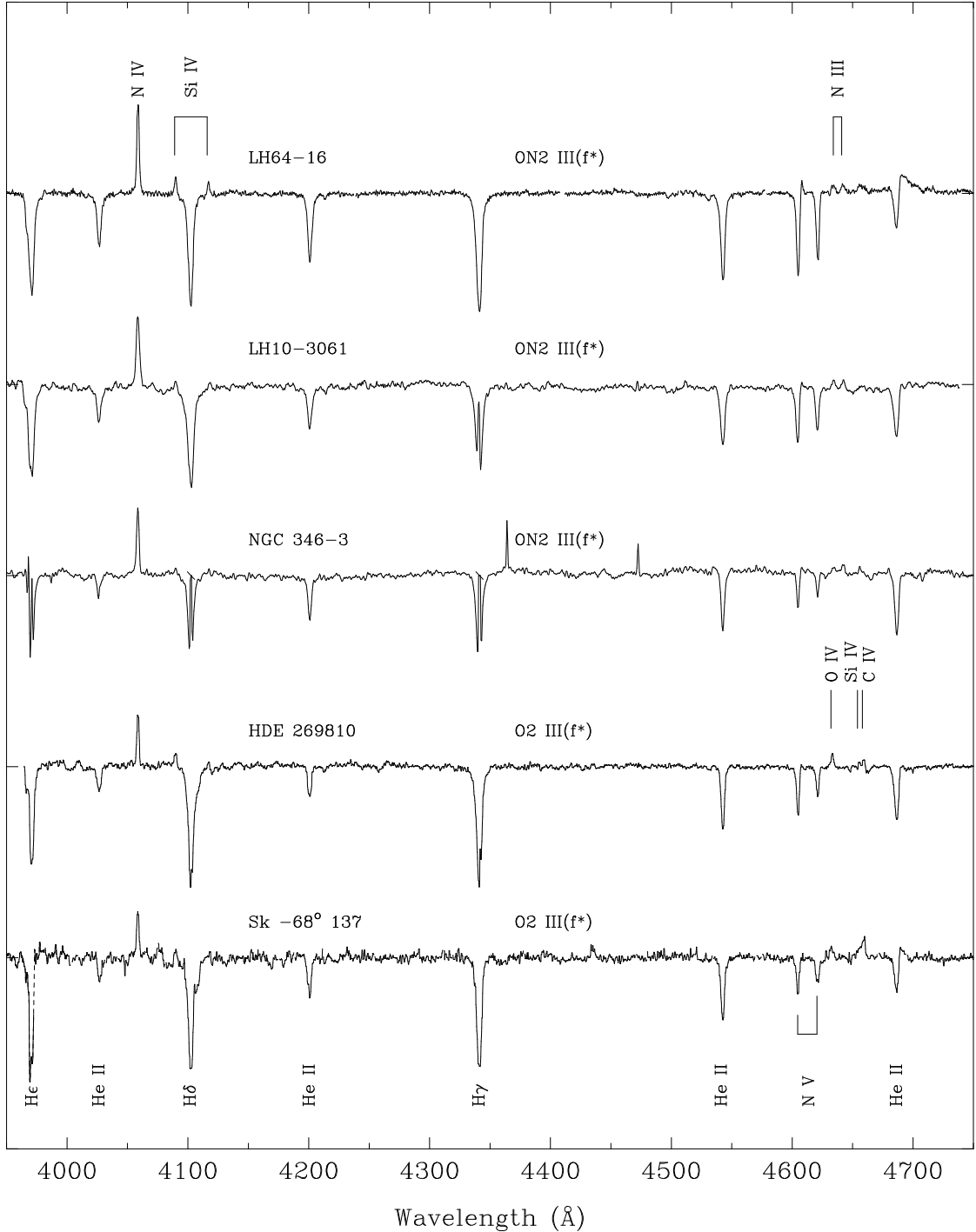}
\caption{
Blue-violet spectrograms of the same stars and with format as in Fig.~1.
The absorption lines identified below are H$\epsilon$ $\lambda$3970,
H$\delta$ $\lambda$4101, and H$\gamma$ $\lambda$4340; He~II 
$\lambda\lambda$4026, 4200, 4541, 4686; and N~V $\lambda\lambda$4604$-$20.
At the top, N~IV $\lambda$4058; Si~IV $\lambda\lambda$4089-4116; and
N~III $\lambda\lambda$4634, 4640-42 emission lines are identified.
Above HDE~269810, O~IV $\lambda$4632, Si~IV $\lambda$4654, and C~IV
$\lambda$4658 emission lines are identified; this O~IV line had been 
identified with Si~IV $\lambda$4631 by Walborn et~al.\ (2002), but we have 
now definitively established from its wavelength and intensity in 
high-resolution data (Fig.~5) that the O~IV contribution is dominant. 
Hydrogen nebular emission lines have been truncated in NGC~346$-$3.  Note 
the different N/C and He/H line ratios between the ON2 and O2 classes 
(the latter as permitted by nebular H~emission contamination).
\label{fig2}}
\end{figure}

\begin{landscape}
\centering
\begin{figure}
\includegraphics[width=.9\textwidth,angle=-90]{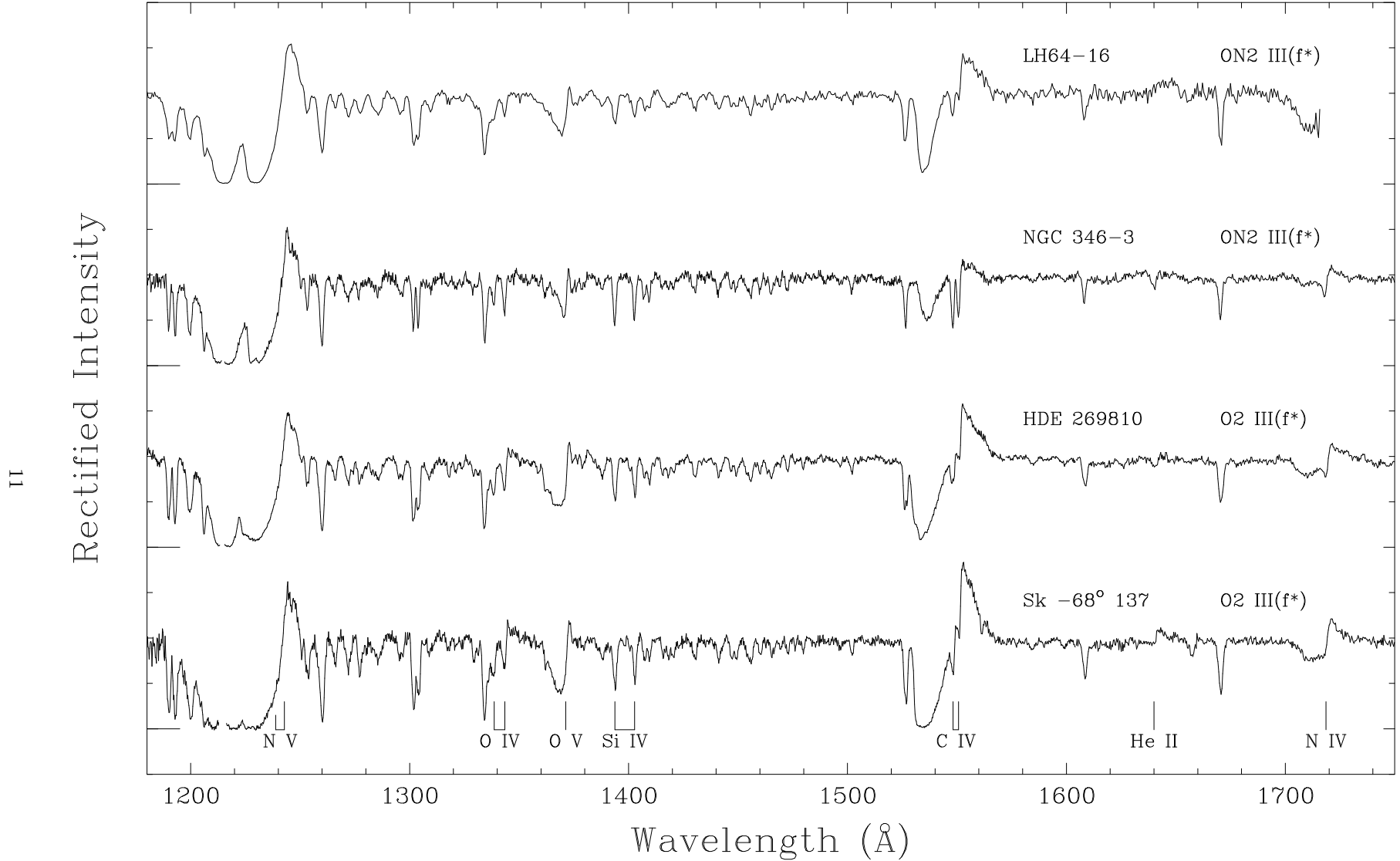}
\caption{
\emph{HST} far-ultraviolet spectrograms of four stars shown in the previous 
figures.  The longer ordinate tick marks give the zero levels and the
continua are normalized to 1.0.  The spectral features identified below
are N~V $\lambda\lambda$1239-43, O~IV $\lambda\lambda$1338$-$43, O~V
$\lambda$1371, Si~IV $\lambda\lambda$1394-1403 (interstellar), C~IV
$\lambda\lambda$1548$-$51, He~II $\lambda$1640, and N~IV $\lambda$1718.
Note the difference between the relative strengths of the N~V and C~IV
features between the upper and lower pairs of spectra; the O~IV and O~V
features also appear weaker in the ON2 spectra.
\label{fig3}}
\end{figure}
\end{landscape}

\begin{figure}
\centering
\includegraphics[width=.85\textwidth]{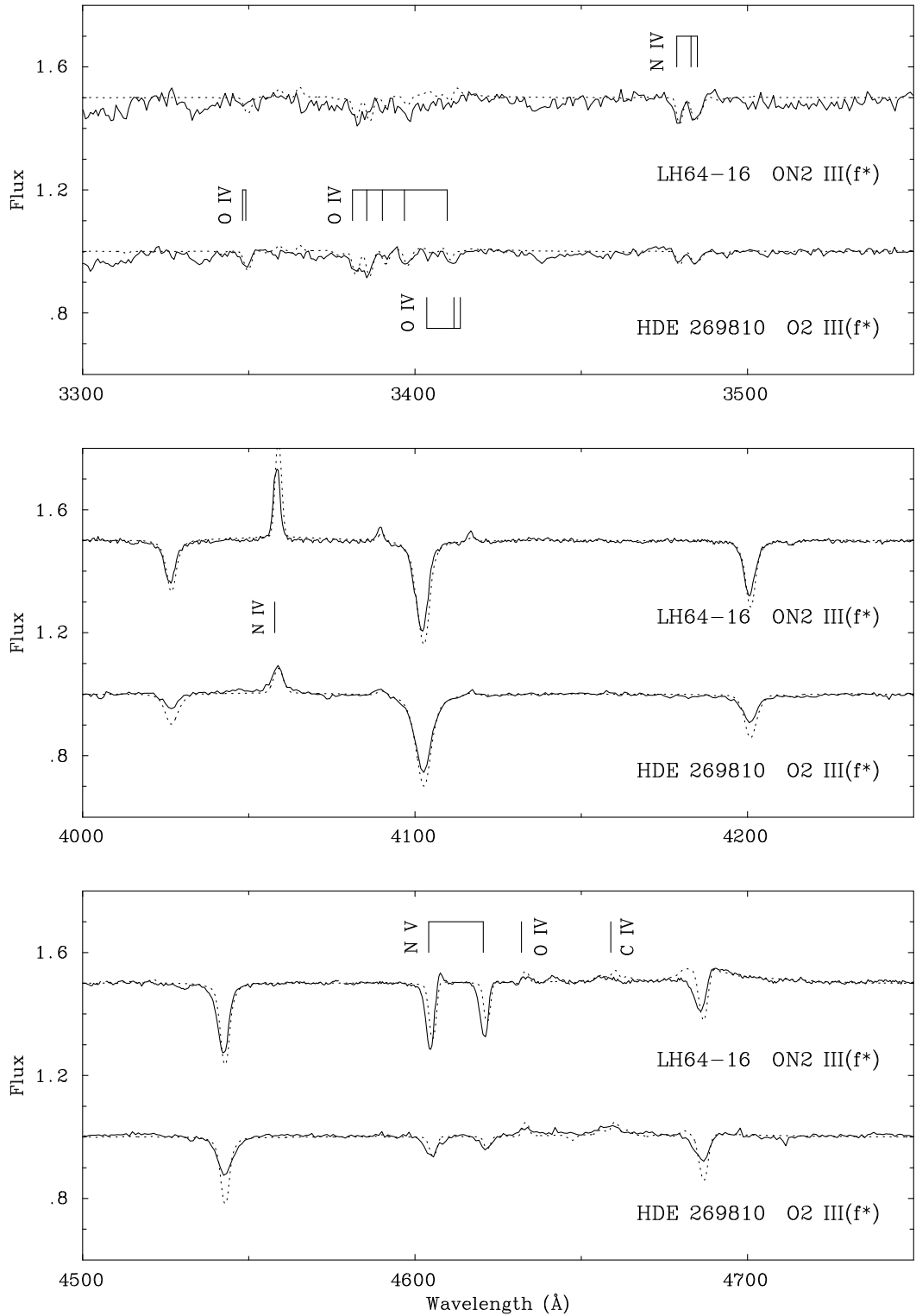}
\caption{
Atmospheric model fits ({\it dotted lines\/}) to key diagnostic features in 
two of the intermediate-resolution spectrograms ({\it solid lines\/}).  The 
56kK fit to LH64$-$16 is shown (see text).  The spectral lines identified in 
the top panel are O~IV $\lambda\lambda$3348$-$49, 3381$-$85$-$90$-$97$-$3410, 3404$-$12$-$14 
and N~IV $\lambda\lambda$3479$-$83$-$85.  In the middle panel, N~IV $\lambda$4058 
is identified, and in the bottom panel, N~V $\lambda\lambda$4604$-$20, 
O~IV $\lambda$4632, and C~IV $\lambda$4658.
\label{fig4}}
\end{figure}

\begin{figure}
\centering
\includegraphics[width=.8\textwidth]{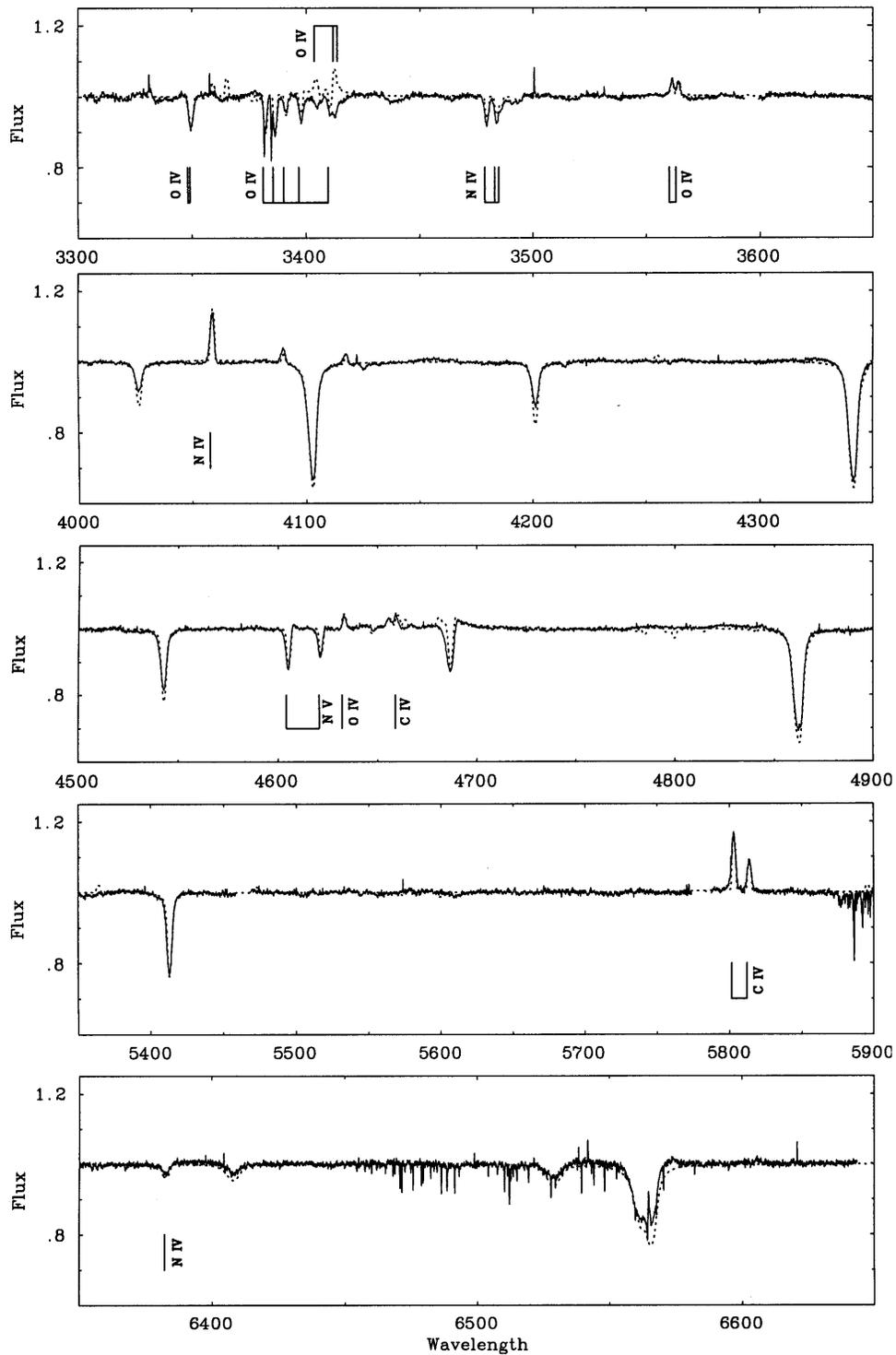}
\caption{
Atmospheric model fit ({\it dotted lines\/}) to the VLT/UVES high-resolution 
observation of HDE~269810 ({\it solid lines\/}).  The spectral lines
identified by panel from top to bottom are O~IV $\lambda\lambda$3348$-$49,
3381$-$85$-$90$-$97$-$3410, 3404$-$12$-$14, 3560$-$63 and N~IV $\lambda\lambda$3479$-$83$-$85; 
N~IV $\lambda$4058; N~V $\lambda\lambda$4604$-$20, O~IV $\lambda$4632, and C~IV 
$\lambda$4658; C~IV $\lambda\lambda$5801$-$12; and N~IV $\lambda$6381.
\label{fig5}}
\end{figure}

\begin{figure}
\centering
\includegraphics[width=.5\textwidth]{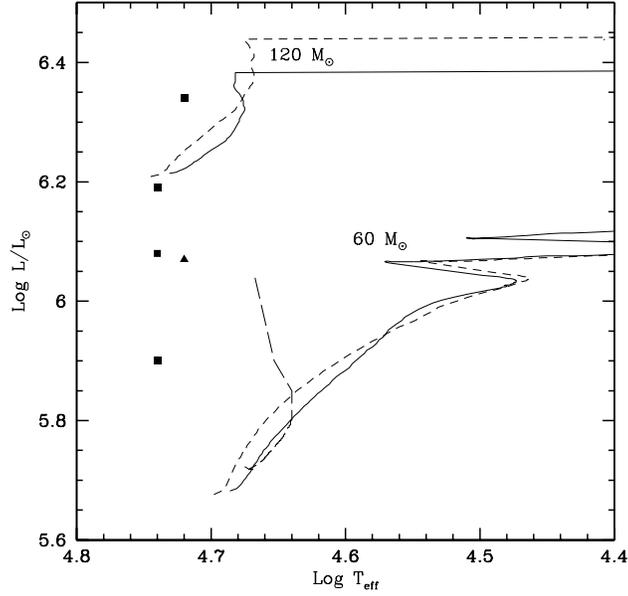}
\caption{
Theoretical HR Diagram comparing the observed stars with Geneva evolutionary 
tracks at both LMC and SMC metallicities, for an initial rotational velocity 
of 300~km~s$^{-1}$.  Also included is a 60~$M_{\odot}$ homogeneous track
at solar metallicity, for an initial rotational velocity of 400~km~s$^{-1}$.  
{\it Squares\/}, LMC stars; {\it triangle\/}, SMC star; {\it solid lines\/}, LMC 
tracks; {\it short dashed lines\/}, SMC tracks; {\it long dashed line\/},
homogeneous track.
\label{fig6}}
\end{figure}

\begin{figure}
\centering
\includegraphics[width=.5\textwidth]{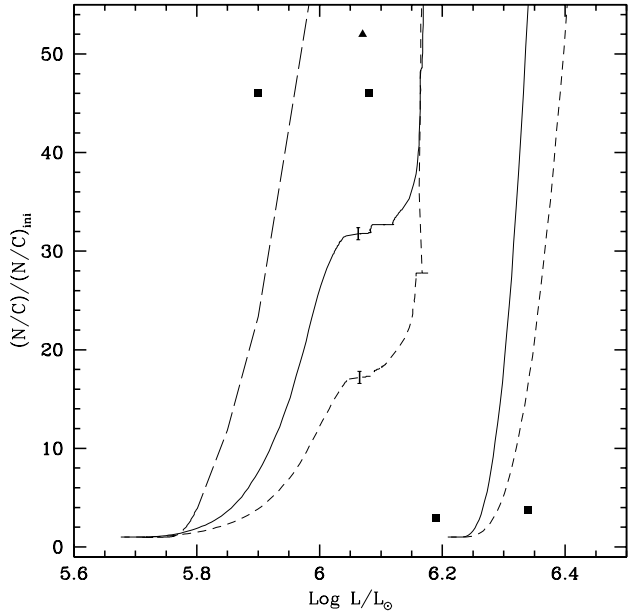}
\caption{
Derived stellar N/C abundance ratios relative to their initial (interstellar) 
values compared with predictions from the models in Fig.~6.  The 60~$M_{\odot}$ 
tracks are toward the left and the 120~$M_{\odot}$ toward the right; short 
vertical line segments on the former indicate the end of the hydrogen-burning 
main sequence.  Other symbols and line styles as in Fig.~6.
\label{fig7}}
\end{figure}

\begin{figure}
\centering
\includegraphics[width=.5\textwidth]{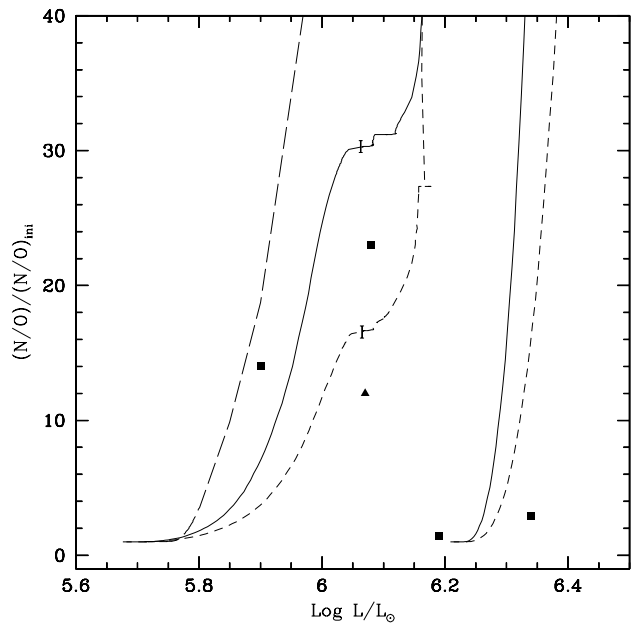}
\caption{
Derived stellar N/O abundance ratios relative to their initial (interstellar)
values compared with predictions from the models in Fig.~6.  Format, symbols, 
and line styles as in Fig.~7.
\label{fig8}}
\end{figure}

\end{document}